\documentclass[12pt,preprint]{aastex}

\newcommand{\simgt} {\,\hbox{\lower0.6ex\hbox{$\sim$}\llap{\raise0.6ex\hbox{$>$}}}\,}
\newcommand{\simlt} {\,\hbox{\lower0.6ex\hbox{$\sim$}\llap{\raise0.6ex\hbox{$<$}}}\,}

\shorttitle{GC subpopulations in major dry mergers}
\shortauthors{Shin \& Kawata}

\begin{document}

\slugcomment{Accepted to ApJ}

\title{THE SPATIAL DISTRIBUTIONS OF RED AND BLUE GLOBULAR 
CLUSTERS IN MAJOR DRY MERGER REMNANTS}

\author{Min-Su Shin}
\affil{Princeton University Observatory, Peyton Hall, Princeton, NJ 08544-1001}
\email{msshin@astro.princeton.edu}

\and

\author{Daisuke Kawata}
\affil{The Observatories of the Carnegie Institution of Washington, 
813 Santa Barbara Street, Pasadena, CA 91101}
\affil{Swinburne University of Technology, Hawthorn VIC 3122, Australia}
\email{dkawata@ociw.edu}

\begin{abstract}
 Using high-resolution N-body simulations, we examine whether 
a major dry merger mitigates the difference in 
the radial density distributions between red and blue 
globular clusters (GCs).
To this end, we study the relation between the density slope of the GCs in 
merger progenitors and that in a merger remnant, 
when the density distribution 
is described by $n_{\rm GC}\propto r^{-\alpha}$. 
We also study how our results depend on the merger orbit 
and the size of the core radius of the initial GC density 
distribution. 
We find that a major dry merger makes the GC profile flatter,
and the steeper initial GC profile leads to more significant flattening,
especially if the initial slope is steeper than $\alpha\sim3.5$. 
Our result suggests that if there is a major dry merger of
elliptical galaxies whose red GCs have a steeper 
radial profile than the blue GCs, as currently observed,
and their slopes are steeper than $\alpha\sim3.5$,
the difference in the slopes between two populations becomes 
smaller after dry mergers. 
Therefore, the observed slopes of red and blue GCs can be a diagnostic
of the importance of dry merger. 
The current observational data show that the red and blue GCs have more 
comparable and shallower slopes in some luminous galaxies, 
which may indicate that they have experienced dry mergers.
\end{abstract}


\section{Introduction}

Globular clusters (GCs) in elliptical galaxies have been intensively 
studied in consideration of explaining 
the formation of both their host elliptical galaxies 
and GCs themselves \citep[see][for a review]{araa06}. 
GCs are attractive as tracers of 
the star formation history of their host galaxies \citep[e.g.][]{yi04,strader06},
because some properties of GC systems are correlated with the properties of
their host galaxies \citep[e.g.,][]{brodie91,djorgovski92}.
It is thought that the formation of GCs 
is triggered by starburst accompanying gas-rich galaxy merging 
\citep[e.g.][]{schweizer87,ashman92} or starburst that might 
happen with multiple dissipational collapses \citep{forbes97}. 
Forming young star clusters are found and they are expected to become 
star clusters like current old 
GCs in local galaxy mergers \citep{schweizer06}. 

An important aspect of GC systems in elliptical galaxies 
is a color bimodality
\citep[e.g.,][]{zepf93,geisler96,gebhardt99,larsen01,peng06}.
It is also found that red GCs
are more centrally concentrated than blue GCs \citep[e.g.][]{forte05,bassino06b,
tamura06}, which could put additional constraints on their formation scenario \citep{bekki02}. 
The radial profile of each GC subpopulation is well described 
by a power-law distribution, especially at the outer radii,
and the red GCs have a steeper slope than the blue GCs.
Although the origin of this color bimodality is still uncertain, 
it is probably closely related to the 
formation history of their host elliptical galaxies 
\citep[e.g.,][]{yoon06,strader07,kundu07}. 
Classically, three scenarios have been proposed to 
explain the color bimodality; major gas-rich mergers, in situ 
formation of multiple dissipational collapses, and dissipationless accretion. An explanation suggested 
by \citet{ashman92} is that red GCs are metal-rich clusters which might be formed 
by gas-rich disk-disk mergers. Therefore, the red GCs might be younger than blue GCs. 
Meanwhile, \citet{forbes97} explain that blue GCs might be formed in the first stage of dissipational 
collapse and red GCs might be formed after the truncation of the blue GC formation. Another explanation 
given by \citet{cote98} includes accretion of blue GCs from small galaxies into the already formed 
red GCs. More recently, \citet{beasley02} demonstrate that the bimodality can
be explained in elliptical galaxy formation based on a hierarchical clustering scenario.

 On the other hand, recent research suggests that in the late stage of 
evolution, early-type galaxies might have experienced dry merging where merger progenitors 
do not have much gas, nor accompany star formation. 
The number density evolution of red galaxies has been discussed in observations of 
COMBO-17 \citep{bell04} and DEEP2 surveys \citep{faber05}, and such studies suggest that
the density change can be understood by the dominance of dry merging after z $\simlt$ 1
\citep[but see also][]{yamada05,cimatti06,bundy07,scarlata07}.
The evolution of galaxy clustering also implies late 
effects on the evolution of massive red galaxies from dry merging \citep{white07}. Moreover, the 
observations show that dry merging does occur \citep{vandokkum05,tran05,rines07}, 
while the observed features of galaxies are well explained in 
cosmological simulations \citep[e.g.][]{kawata06}.
Recent theoretical studies of dry merger simulations of ellipticals show 
that merger remnants maintain their properties on the fundamental 
plane and other scaling relations \citep[e.g.][]{nipoti03,boylan05,robertson06,ciotti07}. 
The dry merging of binary ellipticals also can 
explain the formation of boxy-type ellipticals \citep{naab06}. 

\citet{bekki06a} demonstrate that the observed correlation 
between a spatial distribution of GCs and the total luminosity of ellipticals
can be explained by sequential dissipationless major mergers,
because the radial density profile of GCs progressively 
flatten after each major dry merger.
This study raises the important question of whether or not 
the slopes of the density profiles of red and blue GCs persist 
after major dry merging. 
For example, we now consider that the density profiles of red and blue GCs
in progenitor elliptical galaxies are described by 
$n_{\rm GC, red}\propto r^{-\alpha_{\rm p,red}}$
and $n_{\rm GC, blue}\propto r^{-\alpha_{\rm p,blue}}$, and 
these profiles in a major dry merger remnant
become $n_{\rm GC, red}\propto r^{-\alpha_{\rm r,red}}$
and $n_{\rm GC, blue}\propto r^{-\alpha_{\rm r,blue}}$.
The current observations suggest that $\alpha_{\rm r,red}>\alpha_{\rm r,blue}$.
However, if a dry merger flattens the red GC density profile more than the blue GC density profile, 
$\alpha_{\rm p,red}-\alpha_{\rm r,red}>\alpha_{\rm p,blue}-\alpha_{\rm r,blue}$,
in the remnant galaxy the difference between $\alpha_{\rm r,red}$ and $\alpha_{\rm r, blue}$
becomes smaller. In this case, the observed difference in the slopes 
of the density profiles of red and blue GCs in nearby ellipticals
can be a valuable diagnostic for the importance of the dry merger in the evolution
of ellipticals. To clarify this issue, the question becomes 
how the flattening of GC profiles during dry merging depends on the initial 
distributions of the GCs.

We use numerical simulations of major dry mergers to 
study the dependence of GC distributions in merger remnants on the initial 
distributions in merger progenitors. 
Then, we can compare $\alpha_{\rm p,red}-\alpha_{\rm r,red}$ and $\alpha_{\rm p,blue}-\alpha_{\rm r,blue}$,
for different sets of $\alpha_{\rm p,red}$ and $\alpha_{\rm p,blue}$.
Since major dry mergers between two equal-mass merger progenitors must leave the most significant 
effects on merger remnants, compared with minor mergers, we study only equal-mass mergers in 
this paper. In \S2, we explain details of our dry merger simulations and initial GC distributions. 
The changes in GC spatial distributions due to major mergers 
are shown in \S3. We discuss the implication of our results in \S4 that is followed by conclusion. 

\section{Simulation and parameters}

 Our galaxy merger simulations are carried out with
a Tree N-body code, {\tt GCD+} \citep{kawata03}.
We model a merger progenitor elliptical galaxy 
which consists of a stellar bulge and
a dark matter (DM) halo.
The stellar component is assumed to follow the \citet{hernquist90}
density profile,
\begin{equation}
 \rho_*(r)=\frac{M_*}{2 \pi}\frac{a}{r}\frac{1}{(r+a)^3}.
\end{equation}
We set the total stellar mass ${\rm M_{*} = 10^{11}\ M_{\sun}}$.
Following \citet{boylan06}, we use an empirical relation
from \citet{shen03},
\begin{equation}
 R_{\rm e}=4.16\left(\frac{M_*}{10^{11} M_{\sun}}\right)^{0.56} {\rm kpc}.
\end{equation}
For the Hernquist profile, $R_{\rm e}$ can be described with
$R_{\rm e}=1.8153\ a$.

The initial DM halo is assumed to follow the NFW density profile
\citep{navarro97};
\begin{equation}
  \rho=\frac{3 H_0^2}{8 \pi G} (1+z_0)^3 \frac{\Omega_0}{\Omega(z)}
 \frac{\delta_c}{c x (1+c x)^2},
\end{equation}
where $c$ is a concentration parameter, and $x=r/r_{200}$ is a radius 
normalized by $r_{200}$ which is the radius of a sphere whose
mean interior density is $200\ \rho_{\rm crit}$.
The characteristic density  $\delta_c$ is
described as
\begin{equation}
 \delta_c=\frac{200}{3}\frac{c^3}{\ln(1+c)-c/(1+c)}.
\end{equation}
The radius of $r_{\rm 200}$ is linked with ${\rm M_{200}}$,
the mass within the radius, with
\begin{equation}
 r_{\rm 200}=1.63\times10^{-2} 
 \left(\frac{M_{\rm 200}}{h^{-1} M_{\sun}} \right)^{1/3}
 \left[\frac{\Omega_0}{\Omega(z_0)}\right]^{-1/3} (1+z_0)^{-1}
 h^{-1} {\rm kpc}.
\end{equation}
We assume $c=10$, $z_0=0$, $H_0=70$ km s$^{-1}$ Mpc$^{-1}$.
The final DM density profile takes into account
adiabatic contraction, following the formula
suggested by \citet{blumenthal86}. 
Following \citet{springel99}, we apply an exponential cut-off 
for the dark matter density profile at $r>r_{\rm 200}$. 
We define the total mass of the DM component 
as the DM mass within $2 r_{\rm 200}$, where the density
is low enough. The total mass of the DM component 
is set to be 
${\rm M_{DM,tot} = M_{NFW}} (2 r_{\rm 200}) - {\rm M_*}$, where
${\rm M_{NFW}}(2 r_{\rm 200})$ is calculated from
the exponentially truncated NFW profile with ${\rm M_{200} = 10^{12}\ M_{\sun}}$.
 The DM and stellar components are composed of $10^5$ 
and $10^4$ particles, respectively, so that the particles
for the different components have roughly same mass.
As a result, the particle mass and softening length 
are ${\rm \sim 10^7\ M_{\sun}}$ and 0.58 kpc, respectively. We also 
run one simulation with half number of particles in 
order to verify convergence of our results. 

 The velocity dispersions for both DM and stellar components 
are assumed to be isotropic, 
$\sigma_r=\sigma_{\theta}=\sigma_{\phi}$,
and follow the Jeans equation. Then, the velocity dispersion 
can be written as 
\begin{equation}
 \rho(r)\sigma_r^2(r)=G
 \int^{\infty}_{r} \rho(s) M(<s) \frac{{\rm d}s}{s^2},
\label{eq:veldisp}
\end{equation}
as shown in \citet{mamon05}.
In our simulations, the two merger 
progenitors have exactly the same distributions of DM and star particles 
for simplicity. 

Three different merger orbits are studied in our simulations as given in Table 1. 
From a fixed initial separation of 400 kpc, we have chosen
three parabolic orbits \citep{barnes92}; 
(A) head-on collision; (B) a close encounter with pericentric 
distance of the orbit, $r_{\rm peri}=5$ kpc; and (C) a wider encounter 
with $r_{\rm peri}=50$ kpc. 
All simulations are stopped after 8 Gyrs even though the merger remnants are 
fully relaxed within 300 kpc within 6 Gyrs.

 For simplicity, we assume that the GCs are spherically distributed 
and that the number density follows the following radial profile:
\begin{equation}
 n_{\rm GC}(r) \propto \frac{1}{(r^{2} + r_{c}^{2})^{\alpha_{\rm p}/2}}, 
\label{eq:ngcr}
\end{equation}
where $r_{\rm c}$ is a core radius. 
We include the core, because the observed GC density profiles suggest the existence of the core 
\citep{forbes96}. We also examine how the initial core size 
affects the slope at the outer radii in a dry merger remnant.
Here, instead of putting additional particles to trace the
positions of GCs during the merger, we randomly select particles from 
DM particles, such that 
the selected GC particles follow the number density distribution 
of equation (\ref{eq:ngcr}) as well as velocity dispersion distribution of 
equation (\ref{eq:veldisp}). 
This sampling of GC particles has the advantage 
that we can simulate various different initial distributions 
without running a simulation several times, which enables us to
study the detailed dependence of the flattening on the initial GC profiles. 

\begin{figure}[t]
\plottwo{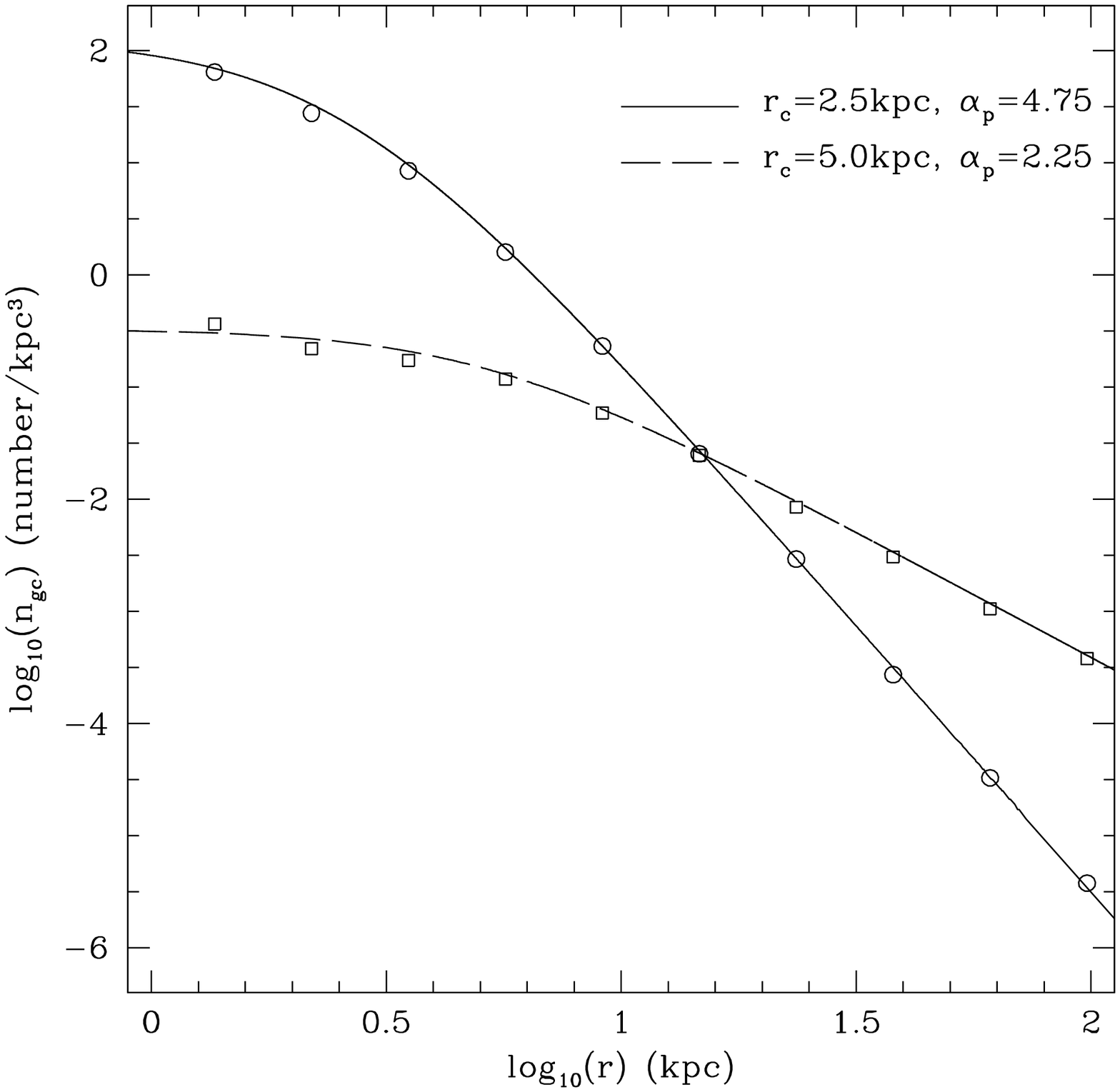}{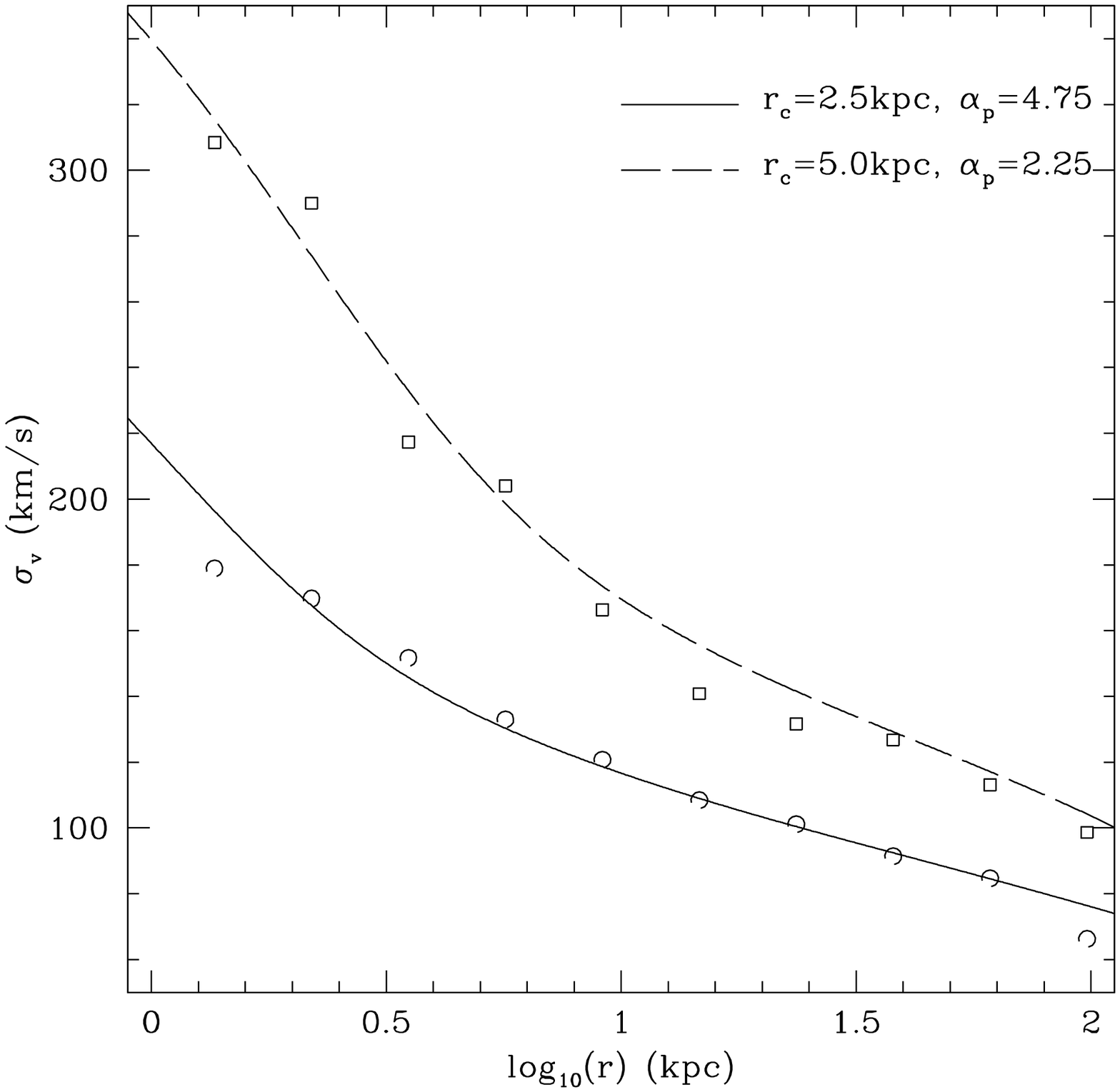}
\caption{Initial number density ({\it left}) and
velocity dispersion ({\it right}) profiles of GCs 
in merger progenitors. The input profiles of $r_{c} = 2.5$ kpc and 
$\alpha_{p}=4.75$ are shown as solid lines, while dashed lines represent 
the profiles of $r_{c} = 5.0$ kpc and $\alpha_{p}=2.25$. 
Dots are the profiles of sampled $10^{4}$ GC particles that 
are well matched to analytical descriptions. 
}
\label{fig:initial_dist}
\end{figure}

 After the merger, we analyze the number density profile
of the selected GC particles.
We are interested in how final slope 
of the GC density profile depends on the initial slope.
We fit the profile of the number density at the radii between 10 and 100 kpc
by  $n_{\rm GC}\propto r^{-\alpha_{\rm r}}$. This radial region is chosen
arbitrarily. However, the important point is that we compare the slope
at a fixed radial region throughout the paper. 
We consider initial density slopes between $\alpha_{\rm p}=2$ and 5, and
the two different core radii of $r_{\rm c}=2.5$ and 5 kpc to see 
the effect of the initial core size on the final slope.
For each initial GC profile, we sample 10,000 GC particles for each merger progenitor.
Figure \ref{fig:initial_dist} demonstrates that our sampling well reproduces
the input number density profile for GCs even at the outer radii. It 
also confirms that the distribution of velocity dispersions for 
the GCs is consistent with the equilibrium values from equation (
\ref{eq:veldisp}). 

\begin{figure}[t]
\plotone{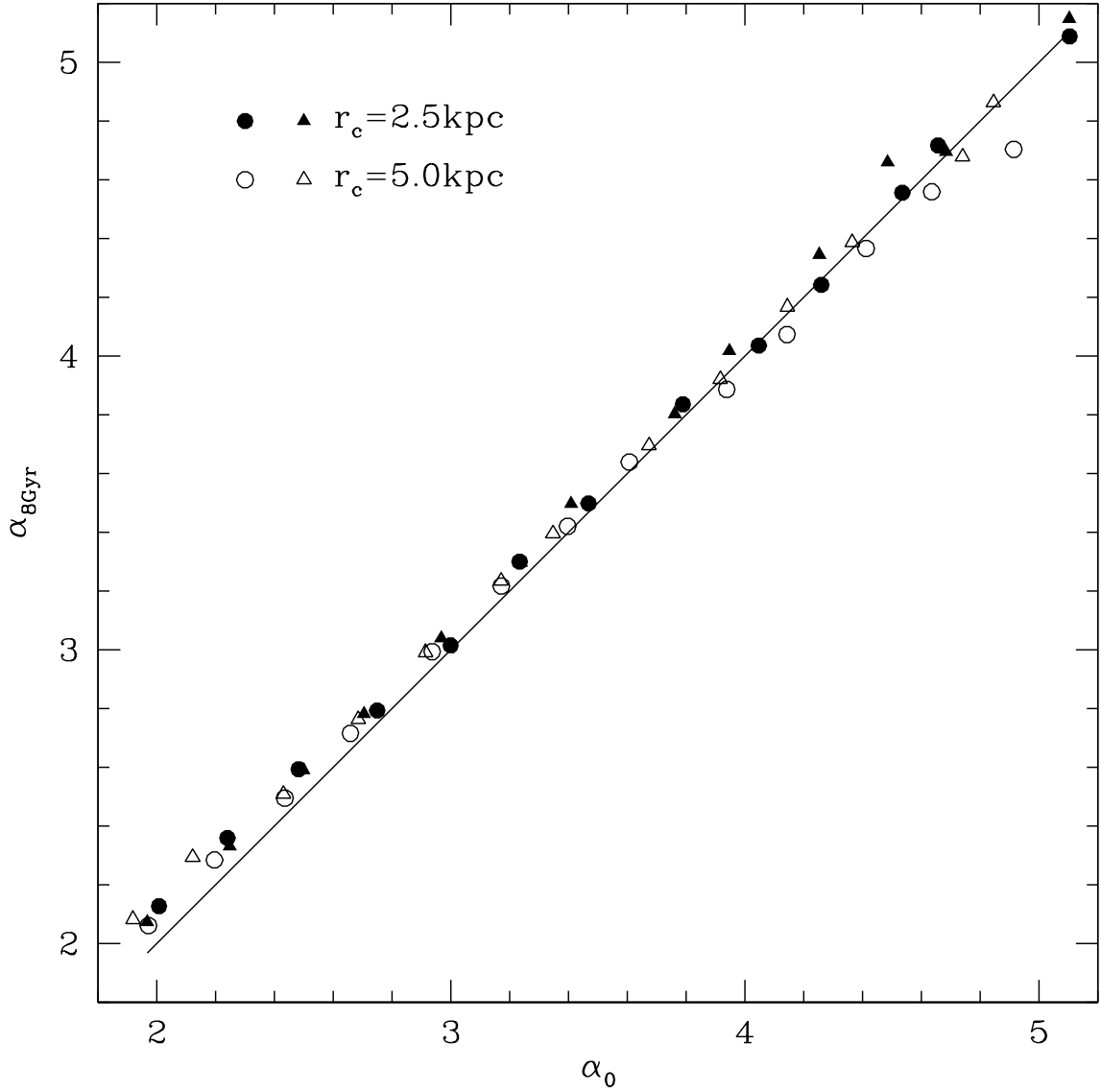}
\caption{Comparison of the initial slope ($\alpha_{0}$)
with the slope ($\alpha_{8Gyr}$) after 8 Gyrs of 
a simulation for an isolated galaxy without any interaction.
Both high-resolution {\it(circle)} and low-resolution {\it(triangle)} 
cases show that the initial distributions of GCs are stable 
in a single isolated galaxy for 8 Gyrs.}
\label{fig:stability}
\end{figure}

 We also test stability of the initial distribution of the selected 
GC particles 
by running a simulation of a single isolated galaxy with the sampled 
GC particles. After 8 Gyrs, the isolated single galaxy maintains 
the initial distribution of the GCs, as shown in Figure 
\ref{fig:stability}.

\section{Results}

\begin{figure}[t]
\plotone{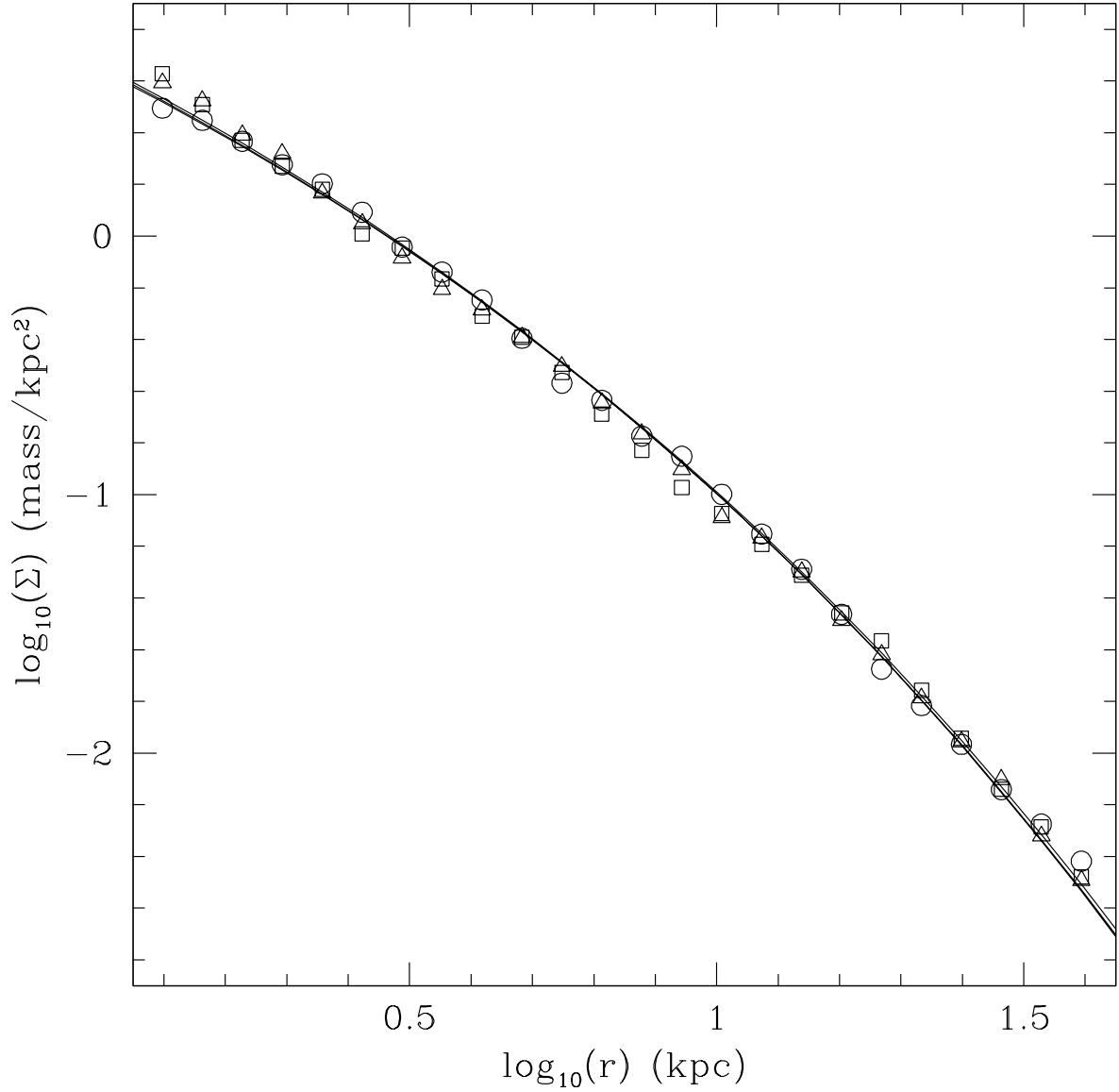}
\caption{Projected mass density distribution of stars in merger remnants. The 
de Vaucouleurs profile {\it(line)} is well matched to stellar distributions in 
all merger remnants. The distributions of the simulation A, B, and C are 
represented by triangle, rectangle, and circle, respectively. The measured 
effective radius is about 6.2 kpc.}
\label{fig:merger_remnant}
\end{figure}

We present the projected density distribution of the stellar component 
of a merger remnant in Figure \ref{fig:merger_remnant}. 
We can see that there is a negligible effect from  
the difference in orbits on the distribution of stars
in the merger remnant. We also fit the final stellar density profiles
between $r=1$ and 40 kpc with de Vaucouleurs profiles. 
In Figure \ref{fig:merger_remnant}, de Vaucouleurs profile describes well 
the projected distribution of the stellar component in all merger remnants 
from the three different orbits. 
The measured effective radius is close to 
$\sim$ 6.2 kpc for all three simulations, which is similar
to the results in \citet{boylan05}. 

\begin{figure*}[t]
\plotone{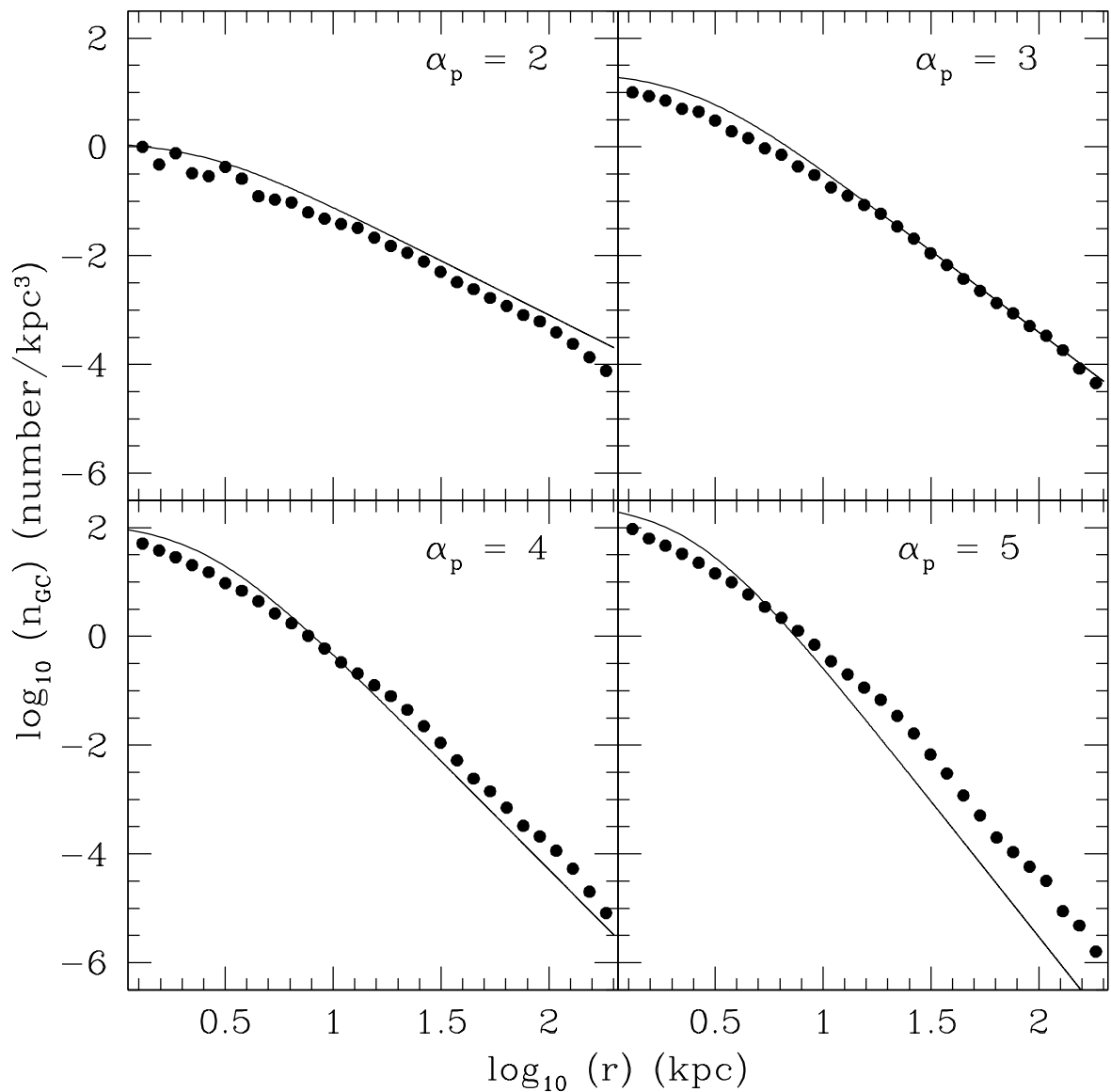}
\caption{GC distributions in merger remnants for simulation B. 
The initial distributions of $r_{c} = 2.5$ kpc are shown for  
$\alpha_{p} = 2.0, 3.0, 4.0$, and 5.0 {\it(solid line)}. 
The distributions of GCs in merger 
remnants are represented by dots. The initial distributions are scaled 
upward by a factor of two in order to match the 
total number of GCs in merger remnants.}
\label{fig:n_dependence}
\end{figure*}

In Figure \ref{fig:n_dependence} we show the final GC number density
profile for the different models with $\alpha_{\rm p}=2$, $4$ 
and $5$ for simulation B. 
We first focus on the results for simulation B,
and discuss the orbital dependence later. 
For models of $\alpha_{\rm p}=4$ and $5$, the GC profile 
in the merger remnants is clearly flatter than the
initial one. This result is consistent with what have been found
in similar dissipationless galaxy merger simulations 
\citep{white78,white80,villumsen82,villumsen83,duncan83,white83,bekki06a}.
On the other hand, for the $\alpha_{\rm p}=2$ model 
there is not much change in the density slope of the GC distribution in a merger remnant. 
In addition, comparing models of 
$\alpha_{\rm p}=4$ and $5$, the difference between
the initial and the final profile is more significant
for the $\alpha_{\rm p}=5$ model. Hence, flattening of the GC
radial profile due to a major dry merger clearly depends on 
the GC density profile in a progenitor galaxy.

 Figure \ref{fig:alpha} summarizes our main results.
We find that a higher $\alpha_{\rm p}$ 
leads to a larger $\alpha_{\rm p}-\alpha_{\rm r}$. In other words,
the steeper initial GC profiles result in stronger flattening. 
The boundary between a significant and a weak flattening is 
around $\alpha_{\rm p} \sim 3.5$. 
In the next sections, we discuss the implication of our results for 
the evolution in the relative distributions of red and blue GCs during 
a major dry merger. First, we briefly examine 
the importance of the merger orbits and the initial core radii.

\subsection{Merger Orbits}

 We find that the final GC density profile
weakly depends on a merger orbit as shown in Figure \ref{fig:orbit_dependence1}. 
The found maximum difference of $\alpha_{\rm r}$ is $\sim 0.2$ for $\alpha_{p} = 5$. 
As also highlighted in Figure \ref{fig:orbit_dependence2} of the three orbits, 
the simulation C causes the least flattening in the GC distribution.
The highest angular momentum collision is not efficient in flattening GC distributions. 

 We also find that the density profile in the outer regions 
of merger remnants are more sensitive to the merger orbits
than that in the inner region. 
In Figure \ref{fig:orbit_dependence2}, all three merger orbits 
do not produce any differences in $r < 10$ kpc for 
an initial $\alpha_{\rm p} = 5$ model.
However, beyond $r \sim 10$ kpc, the small difference of flattening 
begins to appear.

 As a convergence test, we also run simulation A with
a half number of particles, i.e, $5\times10^4$  DM,
$5\times10^3$  stellar, and $5\times10^3$ GC particles.
 Figure \ref{fig:stability} demonstrates
that the low resolution simulation also provides
a stable initial condition.
As shown in Figure \ref{fig:convergence}, the low-resolution simulation
leads to the consistent results to the high-resolution simulation.
Thus, our results are not affected by the number of particles adopted.

\subsection {Dependence on the size of core radius}

 In Figure \ref{fig:r_c_dependence1}, we compare the GC slopes after major
dry mergers in the cases where the merger progenitors
have the core radii of $r_{\rm c} = 2.5$ and 5 kpc. 
Larger initial core radii result in stronger flattening. 
Therefore, it is clear that the flattening due to a dry merger
depends on the initial core radius. 

The size of the initial core radius is a more important factor 
for the final density slope than the merger orbit, 
as found in the comparison between Figure \ref{fig:orbit_dependence1} 
and \ref{fig:r_c_dependence2}. In particular, 
the size of the initial core radius affects the steepening
of the GC density profile 
differently, depending on $\alpha_{\rm p}$.
The difference in $\alpha_{\rm r}$ between the different 
initial core radii is more significant
for a larger $\alpha_{\rm p}$ model than for a smaller $\alpha_{\rm p}$.
Hence it is another interesting issue to see in more detail how 
the initial core radius affects the flattening. 
However, in our method described in \S2,
it is difficult to make the initial distribution of GCs with a very small core. 
Moreover, simulations of a small $r_{\rm c}$ also need higher spatial 
resolution than we currently employ, and so we leave this question for future study.

\section{Discussion and Conclusion}

\begin{figure*}[t]
\plottwo{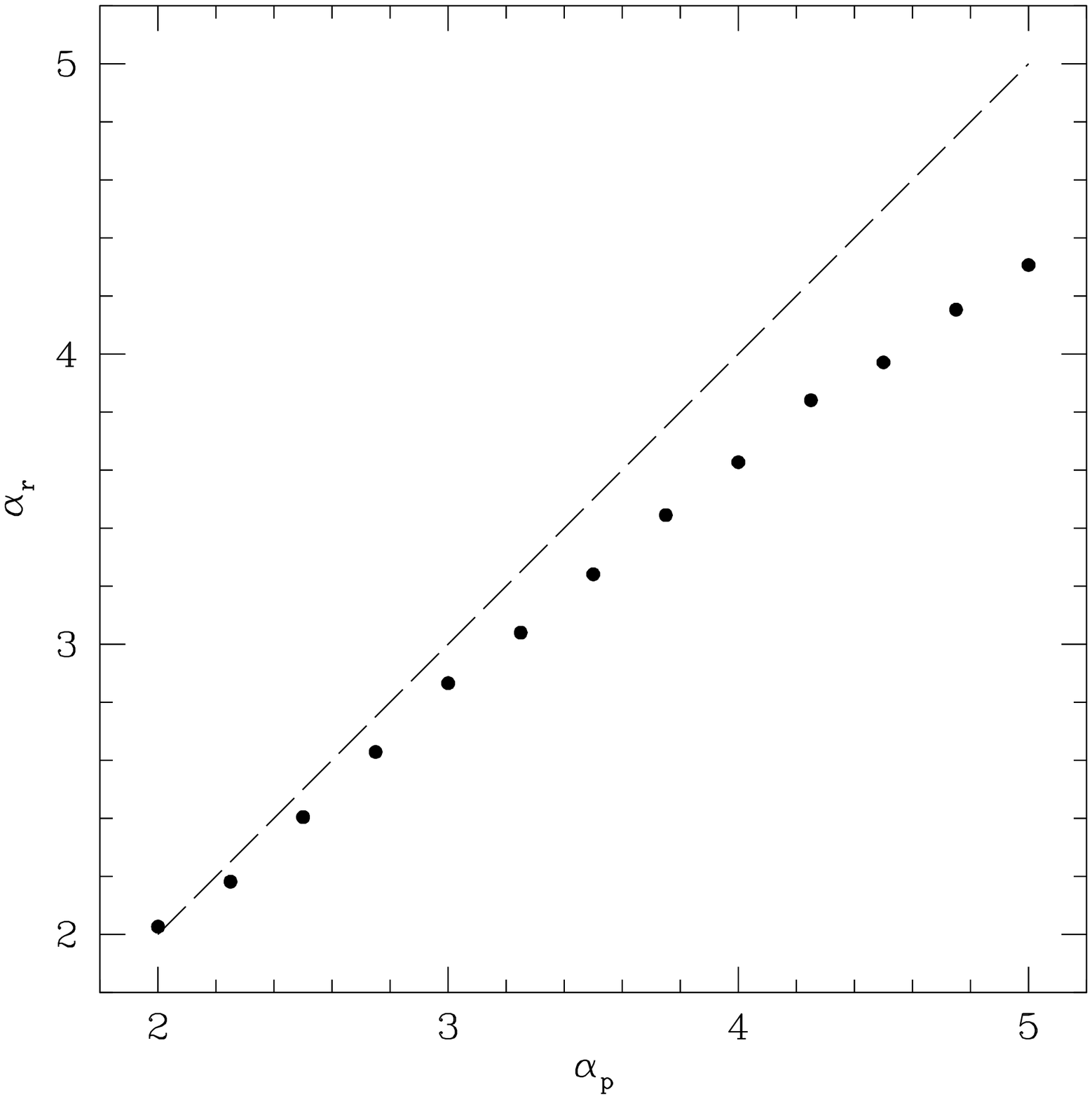}{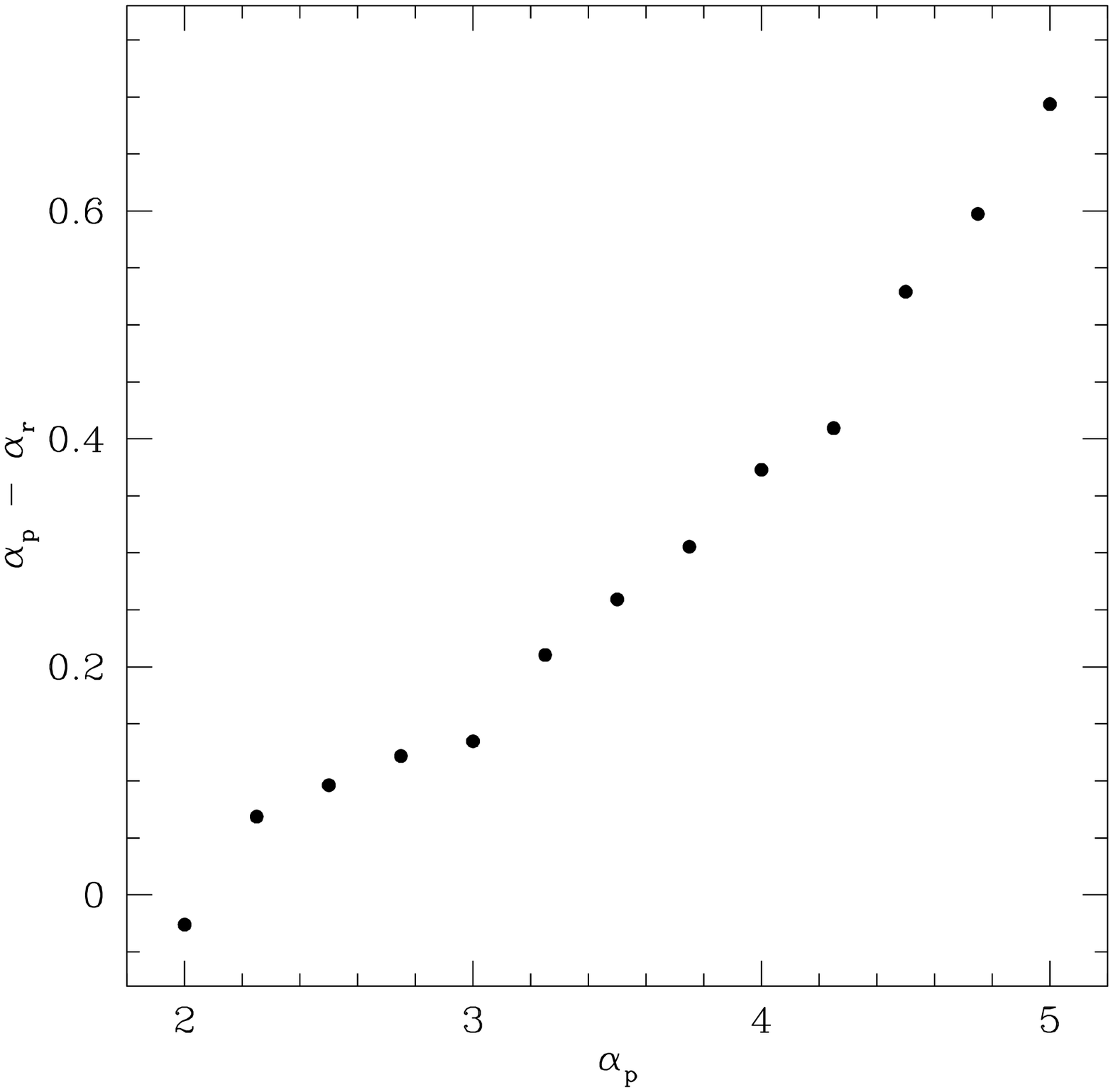}
\caption{$\alpha_{r}${\it (left)} and $\alpha_{p} - \alpha_{r}${\it (right)}  
for simulation B with $r_{c} = 2.5$ kpc. 
The distributions in merger remnants are described as 
$n_{\rm GC} \propto r^{-\alpha_{r}}$ over $10 < r < 100$ kpc. The dashed line represents no 
change in slope.}
\label{fig:alpha}
\end{figure*}

\begin{figure}[t]
\plotone{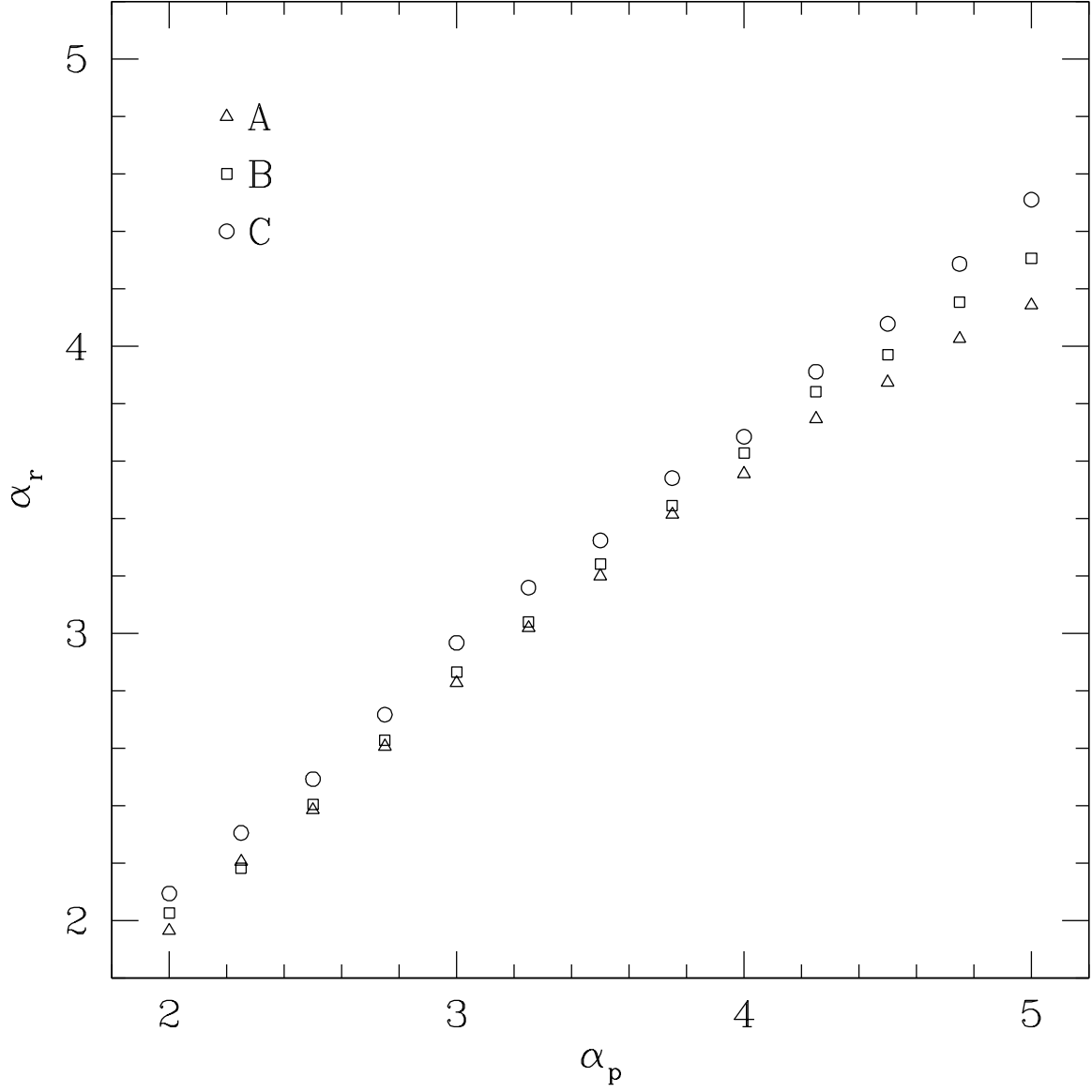}
\caption{Slope of GC distributions in merger remnants produced from three different merger orbits. 
For the initial distribution at a fixed $r_{\rm c}$ = 2.5 kpc, simulation C, with the highest angular momentum,  
shows the smallest change among the three orbits.}
\label{fig:orbit_dependence1}
\end{figure}

\begin{figure}[t]
\plotone{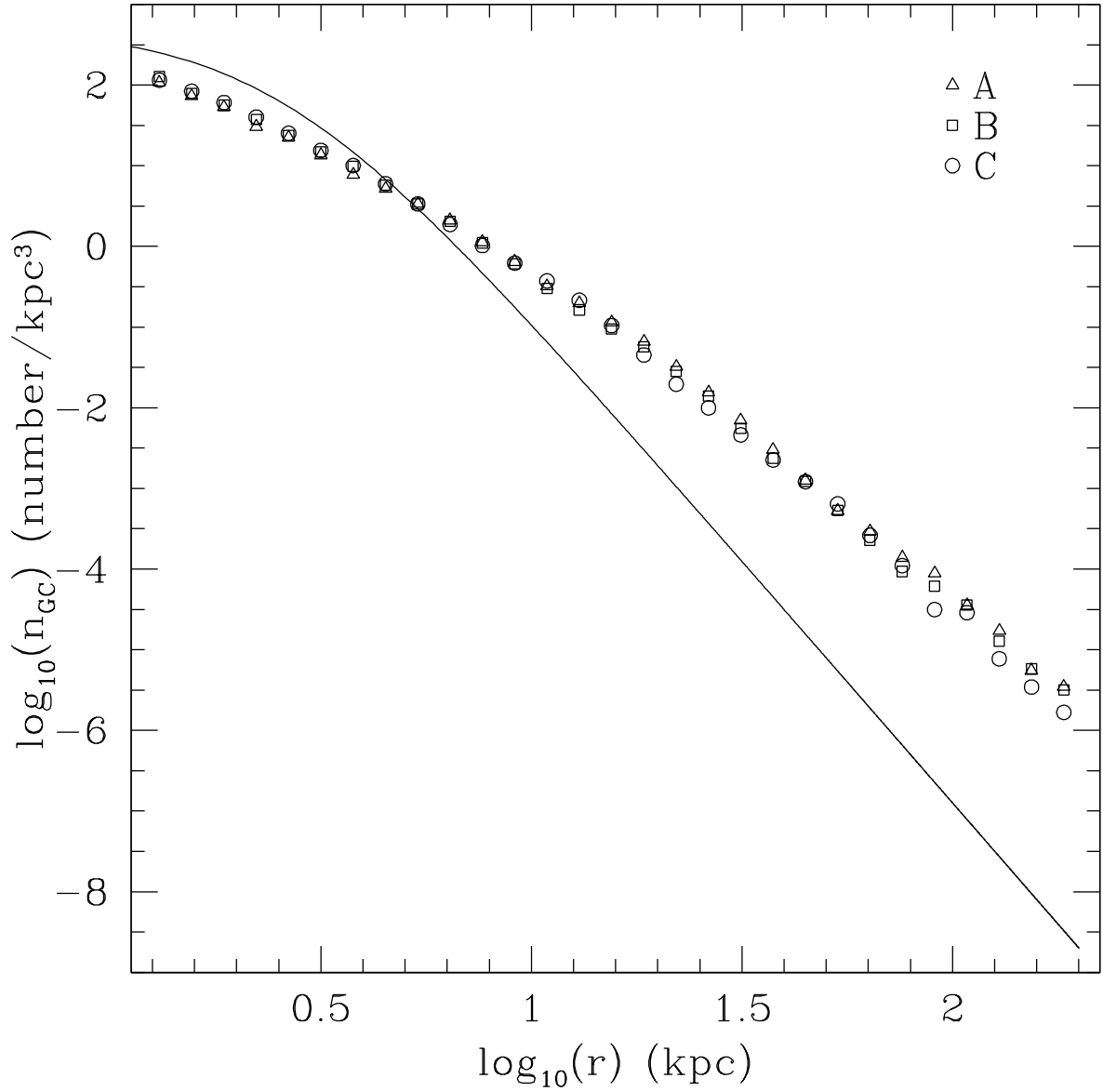}
\caption{Distributions of GCs in three different merger remnants 
for $\alpha_{p} = 5$ and $r_{c} = 2.5$ kpc. 
The effect of different merger orbits 
is stronger in outer region than in inner region. The 
distribution in a merger progenitor is scaled upward by a factor of two for 
case of comparison {\it(line)}.
}
\label{fig:orbit_dependence2}
\end{figure}

\begin{figure}[t]
\plotone{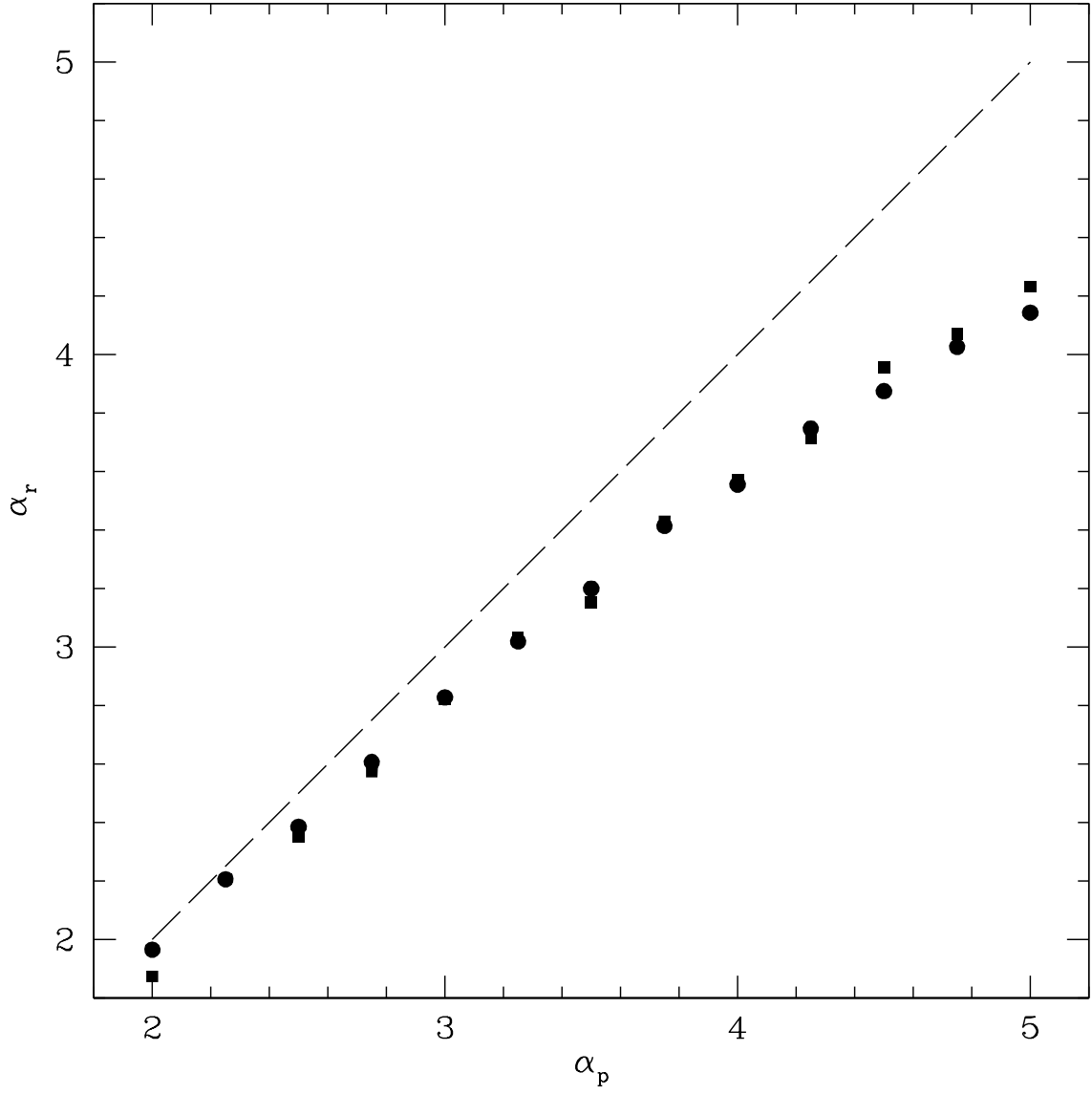}
\caption{
Comparison of the change of slopes ($\alpha_{p}$ vs.\ $\alpha_{r}$)
between the high-resolution {\it(square)}
and low-resolution run {\it(circle)} 
for simulation A with $r_{c}=2.5$ kpc.
The change of slopes in a low-resolution run
is comparable to those of a high-resolution run.
}
\label{fig:convergence}
\end{figure}

 Our main result is 
that steeper initial GC profiles experience stronger flattening 
as shown in Figure \ref{fig:alpha}. 
$\alpha_{\rm p} \approx 3.5$ is in boundary between strong 
and weak flattening. 
The results imply that if the initial slopes 
of both red and blue GCs are steeper than $\alpha_{\rm p} \approx 3.5$,
the difference in the slopes between two populations of GCs
will become much smaller, independent of merger orbits. 
In particular, even only one dry merger can make the slope flatter
dramatically.
Moreover, it also makes both slopes be around $\alpha_{\rm r} \approx 
3.5 - 4.5$. 
For example, if $\alpha_{\rm p,red}=5$ and $\alpha_{\rm p, blue}=4$,
Figure \ref{fig:alpha} suggests that 
$\alpha_{\rm r,red} \approx 4.3$ and $\alpha_{\rm p, blue} \approx 3.6$.
The difference in the slope between red and blue GCs 
becomes significantly small. 
Therefore, the difference in the slopes of red and blue GCs 
can be a constraint of the number of major dry mergers.

 On the other hand, if the initial GC distribution 
has a slope shallower than $\alpha_{\rm p} \approx 3.5$, 
the slope changes very little, and almost no change in the
case of $\alpha_{\rm p} \approx 2$.
Therefore, if the initial distributions of red and blue GCs 
in merger progenitors 
are steeper than $\alpha_{\rm p} = 2$, and the galaxies
experienced a number of major dry mergers, 
it leads the distributions 
of both red and blue GCs to become close to 
$\alpha_{\rm r} \sim 2$ for both red and blue GCs,
and the difference in the slope of spatial distributions becomes 
difficult to be measured.

\begin{figure}[t]
\plotone{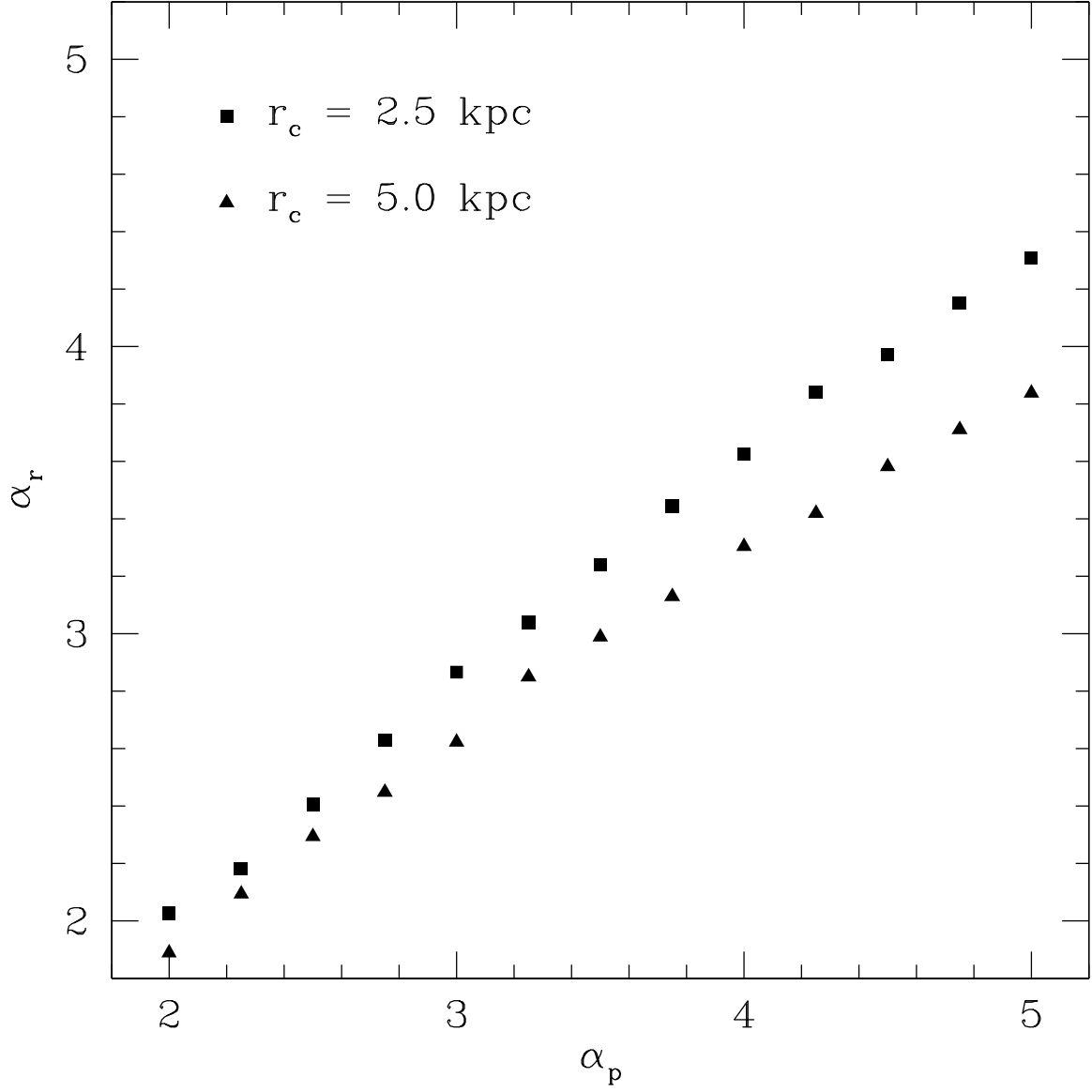}
\caption{
Dependence of GC distributions in merger remnants 
on the initial core radius, $r_{\rm c}$. 
Simulations B for $r_{\rm c} = 2.5$ and 5.0 kpc are compared. 
}
\label{fig:r_c_dependence1}
\end{figure}

\begin{figure}[t]
\plotone{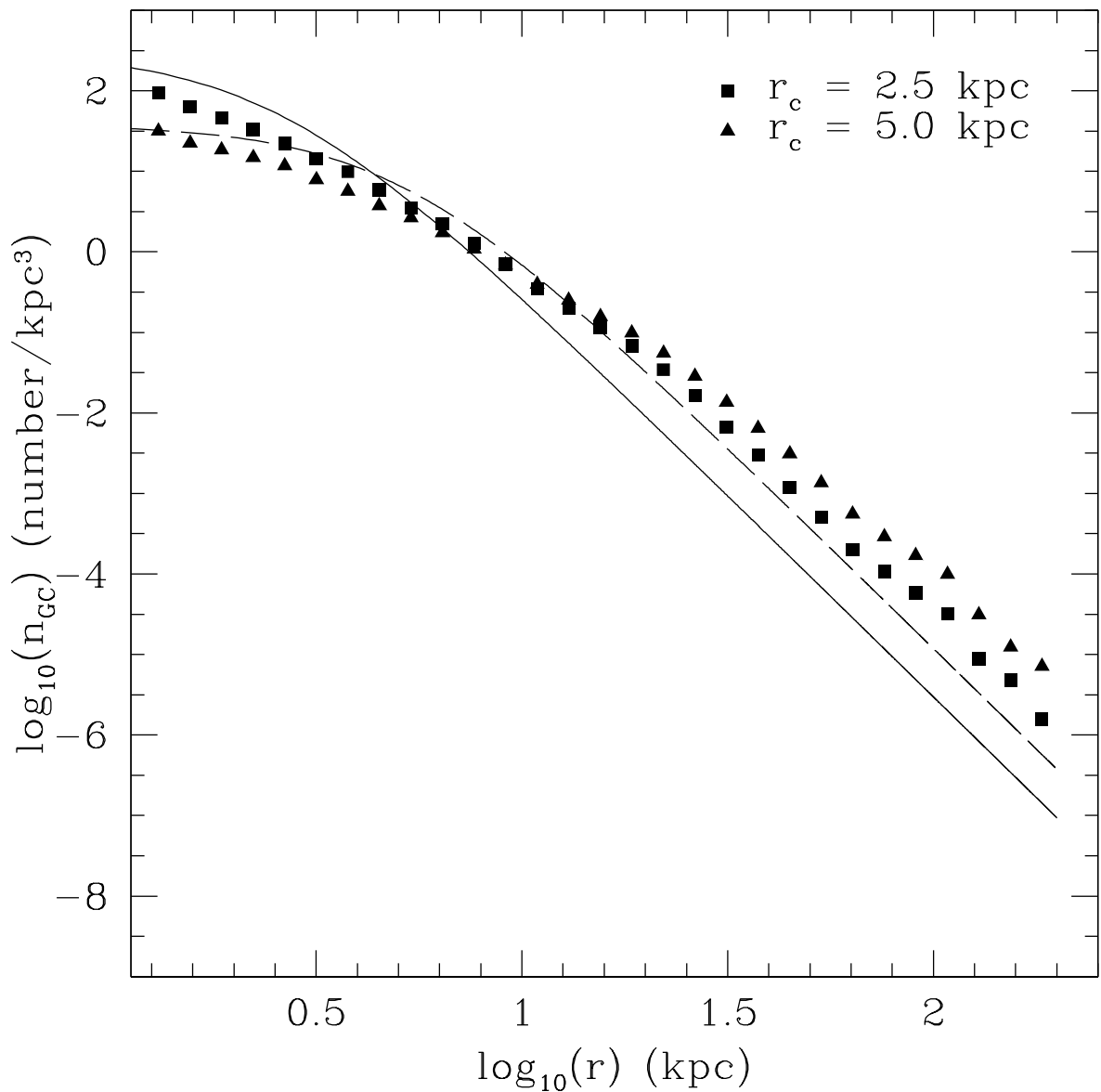}
\caption{Distributions of GCs in merger remnants for $r_{\rm c} = 2.5$ and 5.0 kpc. 
For simulation B and $\alpha_{\rm p} = 5$, we present initial 
distributions in merger 
progenitors for $r_{\rm c} = 2.5$ {\it(solid line)} and 5.0 kpc {\it(dashed line)} and distributions 
in the merger remnant. 
}
\label{fig:r_c_dependence2}
\end{figure}

 It is also worth noting that 
if the initial core size is larger for blue GCs
than for red GCs, but their initial slopes are the same,
a major dry merger can make the slope for the blue GCs
shallower than for the red GCs,
as shown in Figure \ref{fig:r_c_dependence1}. 
Therefore, the final slope is not a simple function of
the initial slope.

 Our results suggest that dry mergers make 
the slopes of the density profiles for red and blue GCs
shallower and similar. 
Several studies \citep[e.g.][]{kissler97,vandenbergh98,forbes05,lauer07,
emsellem07} claim that 
the various observed properties for ellipticals 
show a transition around ${\rm M_{V} \sim -21}$, 
i.e. ${\rm \sim 10^{11}\ M_{\sun}}$. 
Some of these bimodalities have been interpreted in the 
picture of dry merger hypothesis for the growth of 
massive ellipticals \citep[e.g.][]{capetti06}. 
For example, \citet{emsellem07} show that 
slowly rotating ellipticals might be mainly affected by dry mergers. The 
slowly rotating ellipticals are more luminous than 
${\rm M_{B} \approx -20.5}$, while less luminous ellipticals are fast rotators. 
It may indicate that less luminous galaxies have not 
experienced any major dry  mergers. 
Therefore, it is interesting to 
examine the observed distributions of red and blue GCs 
as a function of the stellar mass of ellipticals.
If more luminous galaxies have experienced dry mergers,
our results predict that the slopes for red and blue GCs in
bright galaxies are shallower and similar.
Unfortunately, the current observational samples are not
enough to test it statistically. Below we provide some discussion
based on the current limited measurements.

 NGC 1399 is one of ellipticals whose spatial distributions
of red and blue GCs are well-observed \citep{dirsch03,bassino06b}. 
When assuming ${\rm M/L_{V} = 5}$ and $B-V=0.9$ for an elliptical 
galaxy, as used in \citet{forbes05}, the stellar mass of NGC 1399 
is about ${\rm 5 \times\ 10^{11}\ M_{\odot}}$ 
(${\rm M_{B} = -21.8}$) \citep{araa06}. The derived power-law indices of 
projected radial density profiles are $1.9\pm0.06$ and $1.6\pm0.10$ 
for red and blue GCs, respectively \citep{bassino06b}. 
If the distributions are spherically symmetric, the three-dimensional radial 
distributions have power-law slopes of $\alpha \approx 2.9$ 
and $\approx 2.6$ for red and blue GCs, respectively. 
Note that the real slope might be slightly steeper, 
because the distribution of GCs is not infinite. 
The slope difference between red and blue GC distributions  
is $\Delta \alpha \approx 0.3$ while both GC populations show $\alpha 
\approx 3$. In addition to NGC 1399, NGC 1407 is a brightest group galaxy with 
${\rm M_V=-21.86}$, and has a bimodal color distribution of GCs. 
\citet{forbes06} report that the projected slopes of the red and blue
GC density profiles are respectively $1.50\pm0.05$ and $1.65\pm0.29$, 
i.e., expected three dimensional slopes of $\alpha_{\rm red}=2.50$
$\alpha_{\rm blue}=2.65$, which are statistically the same
as each other.

On the other hand, NGC 1374 and NGC 1379 are less luminous than NGC 1399, 
having ${\rm M_{V} = -20.4}$ and $-20.6$, respectively. 
The red and blue GCs of the 
two galaxies are studied in \citet{bassino06a}. The derived projected
density distributions of red and blue GCs in NGC 1374 have slopes of 3.2 and 2.3. 
Hence, the expected slopes in their three-dimensional density profiles are 
$\alpha_{\rm red}\sim 4.2$ and $\alpha_{\rm blue}\sim3.3$. 
The $\Delta \alpha$ in NGC 1379 is $\sim 0.6$, and the red and blue GCs 
have projected power-law slopes of $\sim 2.9$ and $\sim 2.3$, respectively. 
The low-luminosity galaxies have a larger difference in the 
density slopes of the red and blue GCs, 
and their slopes are steeper than 
the more luminous galaxies discussed above. 
Combining with our results, these four sample suggests
that luminous galaxies, like NGC 1399 and NGC 1407, may have experienced
some dry mergers, while less luminous galaxies, such as
NGC 1374 and NGC 1379, may have not been formed through dry mergers,
which is consistent with a scenario that 
the significance of dry mergers depends on the stellar mass.

 However, we also find some giant galaxy that does not follow
the above trend.
Recently, \citet{tamura06} measured the density profiles of red and blue 
GCs for  giant ellipticals: M87 and NGC 4552. We have fitted the projected 
density profile in Figure 5 of their paper by a power-law profile, and 
the derived power-law index is 2.4 and 1.4 for red and blue GCs in M87 
and 1.8 and 1.2 for red and blue GCs in NGC 4552, respectively. 
Although these giant ellipticals have relatively shallower slopes, 
the difference in slopes of red and blue GCs is significant. 
According to our result, these giant ellipticals are unlikely 
to have experienced significant dry mergers.
Some giants might form without dry mergers, or some other
factor, such as a number of minor mergers, might affect the slopes 
of GCs.

 Again, so far there are not many studies which derive 
the density distributions of both red and blue GCs. 
This problem limits the application of our results to the current 
observational data. 
However, we expect that future observations will 
improve the statistics for 
nearby ellipticals, 
and our results will add an important 
framework to interpret the 
galaxies' evolutionary history. It will be also valuable to compare  
spatial distributions of blue and red GCs with other expected 
properties from dry mergers such as surface brightness profile.

\acknowledgments

  We thank Jeremiah P. Ostriker, James E. Gunn, and 
Patricia S\'anchez-Bl\'azquez for many fruitful discussions. 
We also thank Jenny E. Greene for careful reading of 
the manuscript, and Naoyuki Tamura for the data of 
M87 and NGC 4552. We thank the anonymous referee for helpful comments. 
M.-S.S. wishes to thank the Observatories of the Carnegie Institution 
of Washington for its hospitality. 
M.-S.S. was supported in part by the Korean Science and 
Engineering Foundation Grant KOSEF-2005-215-C00056 funded by the Korean government (MOST). 
This research used computational facilities supported by NSF grant AST-0216105.

\clearpage

\begin{deluxetable}{cccc}
\tablewidth{0pt}
\tablecaption{MERGER ORBITS}
\tablehead{
\colhead{Simulation} & \colhead{$\Delta v$ (km/s)} 
& \colhead{$e$} & \colhead{$r_{peri}$ (kpc)}
}
\startdata
A & (206, 0.0, 0.0) & 1.0 & 0.0 \\
B & (205, 23, 0.0) & 1.0 & 5.0 \\
C & (193, 73, 0.0) & 1.0 & 50
\enddata
\tablecomments{Eccentricity, $e$, and perigee distance, $r_{peri}$, are derived from the reduced two-body 
problem \citep[e.g.][]{khochfar06}}
\end{deluxetable}

\clearpage

\end{document}